\documentstyle[prl,epsf,aps]{revtex}

 \newcommand \be {\begin{equation}}
\newcommand \bea {\begin{eqnarray} \nonumber }
\newcommand \ee {\end{equation}}
\newcommand \eea {\end{eqnarray}}

\input{epsf}
\begin{document}
\twocolumn[\hsize\textwidth\columnwidth\hsize\csname@twocolumnfalse\endcsname
\draft      
\title{A moment based approach to the dynamical solution of the Kuramoto model}
\author{C.J. Perez$^{(*)}$ 
and F. Ritort$^{(**)}$}
\address{(*) Departament de Fisica Fonamental\\Facultat de Fisica\\
Universitat de Barcelona
Diagonal 625, 08028 Barcelona (Spain)\\
(**) Institute of Theoretical Physics\\ University of Amsterdam\\
Valckenierstraat 65\\ 1018 XE Amsterdam (The Netherlands).\\ E-Mail:
conrad@ffn.ub.es,ritort@phys.uva.nl}

\date{\today}
\maketitle

\begin{abstract}
We examine the dynamics of the Kuramoto model with a new analytical
approach. By defining an appropriate set of moments the dynamical
equations can be exactly closed. We discuss some applications of the
formalism like the existence of an effective Hamiltonian for the
dynamics. We also show how this approach can be used to
numerically investigate the dynamical behavior of the model without
finite size effects.
\end{abstract} 

\vfill

\vfill
\newpage

\twocolumn
\vskip.5pc] 
\narrowtext
The study of the dynamical behavior of systems with a very large number
of mutual interacting units is a longly debated subject. It is a topic
with interest in many different interdisciplinar fields. The cooperation
between the members of a population may lead to very rich dynamical
situations ranging from chaos, periodicity, phase locking,
synchronization to self-organized critical states, just to cite a few
\cite{Winf,Rev}.  In the presence of disorder such interaction can be
frustrated and this yields new types of behavior. In the realm of
disordered systems much work has been devoted to the study of models
with relaxational dynamics, for instance spin-glass models
\cite{Mez}. In those cases there exists an Hamiltonian function which
governs the dynamics of the system. A large body of information can be
obtained by using the tools of statistical mechanics.  One of the main
results at equilibrium is that the fluctuation-dissipation theorem is
obeyed. But it is definitely interesting to study the dynamical behavior
of dissipative systems in the presence of external driving forces.

A simple model of this type was proposed by Kuramoto to analyze
synchronization phenomena in populations of weakly nonlinearly coupled
oscillators \cite{Kur}. It has become the subject of extensive studies in
recent years due to its applications to biology, chemistry and physics
\cite{var}. The purpose of this letter is to present a new analytical
approach to the Kuramoto model based on the definition of a suitable
hierarchy of moments. It allows to reproduce previous known results and,
in addition, gives a new insight into the nature of the problem. Here,
we will present the method and consider its potential applications
leaving detailed analysis for future work. Through this formalism it is
possible to analyze some aspects of the model that deserve special
attention. As an example, it has been suggested that, under certain
conditions, it is possible to define a suitable Hamiltonian function
from which it is possible to compute stationary properties of the system
within the usual thermodynamic formalism such as ground states and
universality classes at zero temperature \cite{vh} and equilibrium
Boltzmann distribution in the more general case at finite temperature
\cite{cor}. Our method can answer this question in a simple way. We
will also show how our approach can be used to numerically investigate
the behavior of the Kuramoto model free of finite size
effects. Particular results will be obtained for the bimodal
distribution case.

Our formalism complements other recent theories developed to analyze the
Kuramoto model. In particular, it is worthwhile to mention the order
function approach \cite{Dai} useful for studying properties of the
stationary states of the system as well as the critical exponent of the
order parameter at the onset of entrainment. Another interesting method
was proposed in \cite{Craw} based on kinetic theory and suitable to deal
with questions related to the time dependence of the probability density
of the system.

The Kuramoto model is defined by a set of $N$ oscillators whose state can
be specified in terms of only one degree of freedom, the phase. Each phase
$\lbrace \phi_i;1\le i\le N \rbrace$ follows the dynamical equation

\be
\frac{\partial\phi_i}{\partial t}=\omega_i-
\frac{K_0}{N}\sum_{j=1}^N\sin(\phi_i-\phi_j)+\eta_i
\label{eq1}
\ee

\noindent where $\omega_i$ is the intrinsic frequency of the oscillator
randomly chosen from a distribution of density $g(\omega)$, $K_0$ is the
strength of the coupling which, as in the original case, we will consider
ferromagnetic although more complex situations have been analyzed in the
literature \cite{mas}. Finally, $\eta_i(t)$ denotes a gaussian independent
white-noise process 

\be
<\eta_i(t) \eta_j(t')> = 2T \delta_{ij} \delta(t-t').
\label{eqq}
\ee

\noindent Without any other ingredient there is a competition between
the coupling, which tends to synchronize all the oscillators, and the
noise (frequencies plus thermal noise) which breaks the coherence. For a
critical $K_c$ there is a spontaneous transition from incoherence to a
new state where a macroscopic number of units are synchronized.

\noindent To solve the dynamics of the Kuramoto model we define the
the following set of moments,

\be
H_{k}^m=\frac{1}{N}\sum_{j=1}^{N}\,\overline{<\exp(ik\phi_j)>\omega_j^m}
\label{eq2}
\ee

\noindent where $i$ stands for the imaginary unit and $k,m$ are integers
in the range $(-\infty,\infty),[0,\infty)$ respectively. The averages
$<(\cdot)>$ and $\overline{(\cdot)}$ indicate averages over the noise
and frequencies distribution respectively. The definition of this set of
moments is the basis of the new dynamical approach we are proposing. Note
that $H_{k}^m$ is the more natural object we can construct which is
invariant under the local transformation $\phi_i\to\phi_i+2\pi$. This is
also the local symmetry of the dynamical equations (\ref{eq1}). It is
possible to show that the moments $H_{k}^m$ are self-averaging with
respect to the thermal noise
\footnote{The derivation of this result comes from the fact that a
probability density of oscillators can be defined
for the Kuramoto model, see Strogatz and Mirollo in \cite{var}.}.  The
equation of motion for $H_{k}^m$ can be easily derived using the
eq.(\ref{eq2}),

\be
\frac{\partial H_{k}^m}{\partial t}=-\frac{K_0 k}{2}(H_{k+1}^m h_{-1}-
H_{k-1}^m h_1)-k^2 T H_{k}^m\,+\,ik H_{k}^{m+1} 
\label{eq3}
\ee

\noindent where we have defined the $h_k=H_k^0$. The term $k^2 T
H_{k}^m$ can be simply obtained using the Gaussian representation for
the noise $\eta_i$, doing an integration by parts and using the
regularisation condition $\frac{\partial \phi(t)}{\partial
\eta(t)}=1/2$. In this form the equations are closed because the time
operator $\frac{\partial}{\partial t}$ acting on the moment $H_{k}^m$
generates always new moments of the same type. If we define the time
dependent generating function
$g_t(x,y)=\sum_{k=-\infty}^{\infty}\sum_{m=0}^{\infty}
\exp(-ikx)\frac{y^m}{m!}H_k^m(t)$  it is easy to check from eq.(\ref{eq3})
that it satisfies the following differential equation,

\be
\frac{\partial g_t}{\partial t} =-\frac{\partial}{\partial x}
\Bigl (A(x,t)g_t\Bigr )+T\frac{\partial^2 g_t}{\partial
x^2}-\frac{\partial^2 g_t}{\partial x\partial y}
\label{eq4}
\ee
where 

\bea
A(x,t)=-\frac{K_0}{2i}(\exp(-ix)h_{1}-\exp(ix)h_{-1})=\nonumber\\
\frac{K_0}{N}\sum_{j=1}^N\sin(x-\phi_j)=K_0 r \sin(\theta-x)
\label{eq5}
\eea and we have expressed $h_1=h_{-1}^*=r\exp(i\theta)$ where $r$ is
the parameter which measures the coherence (synchronisation) between
oscillators. By substituting eq.(\ref{eq2}) in the definition of the
generating function it is straightforward to check that
$g_t(x,y)=\frac{1}{N}\sum_{j=1}^N\delta(\phi_j-x)\exp(y\omega_j)$.  For
$y=0$ this is nothing else than the probability density of one
oscillator to have phase $x$. In this way we recover the results
obtained by Bonilla \cite{bon87} using the path integral formalism. The
hierarchy of equations (\ref{eq3}) only depends on the time evolution of
the moment $h_1$($h_{-1}=h_1^*$) which is the order parameter of the
problem. The full set of moments are self-consistently computed using
the conditions $h_1(t)=\frac{1}{2\pi}\int_{0}^{2\pi}\exp(ix)g_t(x,0)dx$
and $H_0^m=\overline{\omega^m}$.  In this way, we have reached a
dynamical solution of the problem identical to that found in some
mean-field glassy models where dynamical equations can be exactly closed
\cite{BG}.

\noindent Once we have presented the formulation we present its
applications. The method furnishes a transparent way to show why a
static description based on conventional equilibrium statistical
mechanics cannot give reliable information about the long-time
properties of the Kuramoto model (like the existence of stationary
states). Let us stress from the beginning that our approach deals with
the dynamics of the model once the thermodynamic limit $N\to\infty$ is
taken first than the limit $t\to\infty$. Our results are meaningful in
this case. In the other dynamical approach one does the thermodynamic
limit $N\to\infty$ after solving the dynamics in the long-time limit
$t\to\infty$. This yields quite different results. These considerations
interest specially to the results reported in \cite{vh} where the last
situation has been analyzed. Which of the two approaches is valid relies
on the timescales one is able to observe. Because the crossover time
which matches the two approaches grows extremely fast with the system
size, our approach (which , on the other hand, is the most conventional
one) is more suited to investigate what can be observed in macroscopic
systems in realistic timescales.

Let us consider the following Hamiltonian function,

\be
{\cal H}_{eff}=-\frac{K_0}{N}\sum_{i<j} \cos(\phi_i-\phi_j)
-\sum_i\omega_i\phi_i
\label{eq6}
\ee
where the phases $\phi_i$ are restricted to the interval $[-\pi,\pi)$ in
order ${\cal H}_{eff}$ to be bounded from below. This Hamiltonian function
is the only candidate which generates the equations (\ref{eq1}) in the
Langevin dynamics $\dot{\phi}_i=-\frac{\partial {\cal H}_{eff}}{\partial
\phi_i}+\eta_i$.  We will show that the equilibrium solutions of
(\ref{eq6}) are not stationary solutions of the dynamics eq.(\ref{eq1})
even at zero temperature. To prove this result we compute the partition
function of eq.(\ref{eq6}) at temperature $T=\frac{1}{\beta}$ and evaluate
the moments $H_k^m(eq)$ in equilibrium.  The computations are quite simple
since the disorder in 
${\cal H}_{eff}$ is only site dependent. For sake of simplicity we will
consider here the case in which the disorder distribution is symmetric
$(g(\omega)=g(-\omega))$. It is easy to obtain the equilibrium values of
the different moments. We obtain $H_k^m(eq)=F_k^m(eq)\exp(ik\theta)$
where $\theta$ is an arbitrary phase \footnote{${\cal H}_{eff}$ only
changes by a a constant $\Phi\sum_i{\omega_i}$ if all the phases
$\phi_i$ are changed to $\phi_i+\Phi$} and the $F_k^m(eq)$ are given by

\be
F_k^m(eq)=\overline{\omega^m\,R_k^{\omega}(\beta\,K_0\,r)} 
\label{eq7}
\ee
where the $\overline{A(\omega)}=\int d\omega g(\omega)A(\omega)$ is the
average over the frequency distribution and 
$R_k^{\omega}(x)=\frac{J_k^{\omega}(x)}{J_0^{\omega}(x)}$. The
$J_k^{\omega}(x)$ are generalized Bessel-like functions defined by,

\be
J_k^{\omega}(x)=\int_{0}^{2\pi}\,d\phi\,
\exp(ik\phi+x\,\cos(\phi)+\beta\omega\phi)
\label{eq8}
\ee
where $r$ is fixed by the condition
$r=F_1^0(eq)=\overline{R_1^{\omega}(\beta\,K_0\,r)}$ .

To check if the $F_k^m(eq)$ are stationary solutions of the dynamics we
plug them in eq.(\ref{eq3}) and use the recursion relation,

\bea
J_{k+1}^{\omega}(x)=J_{k-1}^{\omega}(x)-\frac{2(k-i\beta\omega)}{x}J_{k}^{\omega}(x)
\nonumber\\
-\frac{2i\exp(x)}{x}\Bigl (\exp(2\pi\beta\omega)-1\Bigr )~~~
\label{eq9}
\eea
obtaining after some manipulations, 
\be
\frac{\partial F_k^m(eq)}{\partial t}=ikv_m,
\label{eq10a}
\ee
where
\be 
v_m=T\exp(\beta\,K_0\,r)
\overline{\frac{\omega^m \Bigl (\exp(2\pi\beta\omega)-1\Bigr
)}{J_0^{\omega}(\beta\,K_0\,r)}}~~~~~.
\label{eq10}
\ee

Due to the
symmetry property of the $g(\omega)$ it is easy to check that the $v_m$
always vanishes for $m$ even. But it never does for odd $m$. In the
high temperature regime $\beta\to 0$ it is possible to show that
$v_{2m-1}=\overline{\omega^{2m}}+O(\beta^3)$.
In the limit $T\to 0$ it is easy to check that
$v_{2m-1}=\overline{\omega^{2m}}+O(\frac{1}{\beta})$. The general result
is that odd $m$-moments violate stationarity \footnote{An exception to
this rule are the moments $H_0^m$, among them the average frequency,
because $k=0$ and therefore the right hand side of eq.(\ref{eq10a})
vanishes.}.  We conclude that the time derivative of the equilibrium
moments $H_k^m(eq)$ with odd $m$ get an imaginary contribution
proportional to $i=\exp(i\frac{\pi}{2})$ which is transverse in the
complex plane to the $H_k^m(eq)$ itself. Note that the term $v_m$ is the
angular velocity or time derivative of the global phase for all the
moments which only depends on the number $m$. In the case of the bimodal
distribution
$g(\omega)=\frac{1}{2}(\delta(\omega-\omega_0)+\delta(\omega+\omega_0))$
all moments reduce to two different moments (see below) depending if $m$
is even or odd.  In figure 1 we show the $v_{odd}$ as a function of the
temperature in the bimodal distribution for different values of $K_0$
\footnote{In the rest of the paper and without loss of generality we
will take $\omega_0=1$}.  This proves that equilibrium states of ${\cal
H}_{eff}$ at finite temperature and also the ground states of ${\cal
H}_{eff}$ are not stationary states of the dynamics. Note that we cannot
discard the fact that local minima (but not global) of ${\cal H}_{eff}$
at zero temperature are fixed points of the dynamics.

\begin{figure}
\centerline{\epsfxsize=8cm\epsffile{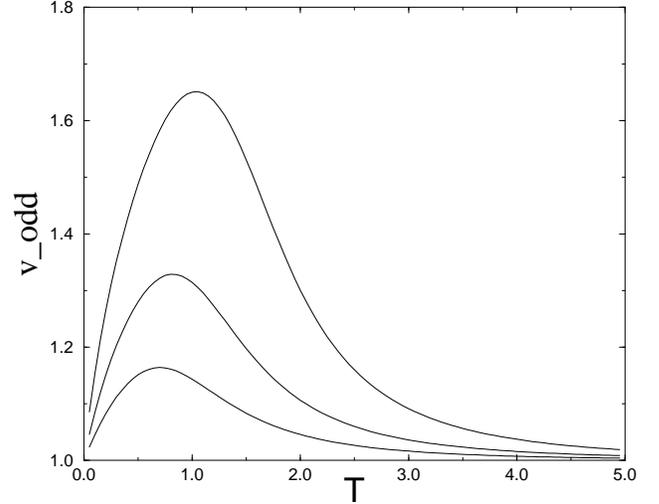}}

\caption{$v_{odd}$ as a function of $T$ for the bimodal distribution for
three values of $K_0=0.5,1,2$ from bottom to top. The cusp in $v_{odd}$
moves to higher temperatures as $K_0$ increases.}
\end{figure}

\noindent Now we want to show how equations (\ref{eq3}) can be used as a
powerful tool to investigate the dynamical behavior of the Kuramoto
model. We will consider the case of the bimodal distribution because the
phase diagram of the model is very rich. We have checked the situation
for other distribution of frequencies and the results have been always
very satisfactory. In this case the full set of moments $H_k^m$ reduces
to two different sets, $f_k=H_k^m$ for $m$ even and $g_k=H_k^m$ for $m$
odd. The full set of dynamical equations read in this case,

\be
\frac{\partial f_k}{\partial t}=-\frac{K_0 k}{2}(f_{k+1} f_{-1}-
f_{k-1} f_1)-k^2 T f_{k}\,+\,ik g_{k}
\label{eq11a}
\ee

\be
\frac{\partial g_k}{\partial t}=-\frac{K_0 k}{2}(g_{k+1} f_{-1}-
g_{k-1} f_1)-k^2 T g_{k}\,+\,ik f_{k}
\label{eq11b} 
\ee

By defining
$\rho_{+}(x)=\frac{1}{2}\sum_{k=-\infty}^{\infty},\exp(-ikx)(f_k+g_k)~~$
and
$\rho_{-}(x)=\frac{1}{2}\sum_{k=-\infty}^{\infty}\,\exp(-ikx)(f_k-g_k)$
we observe that these are the probability densities of having one
oscillator with phase $x$ and natural frequencies $+1$ and $-1$
respectively. By adding and substracting both equations
(\ref{eq11a}),(\ref{eq11b}) we obtain dynamical equations for two sets
of dynamical hierarchies, each set characterized by a population of
oscillators with a given natural frequency (+1 or -1). Note that the two
sets of oscillators are also coupled one to each other through the terms
$ik g_{k}$ and $ik f_{k}$ in (\ref{eq11a}),(\ref{eq11b}).  These
equations also can be used to analytically compute stationary states and
perform stability analysis as has been done in \cite{Bns}. In this last
case it can be shown that the fundamental model $k=1$ decouples form the
rest of the modes in a natural way. Here we follow a different strategy
and use the method to numerically solve the set of equations for a given
number $2L+1$ of terms in the hierarchy $\lbrace f_k,g_k;-L\leq k\leq
L\rbrace$. We stress that $L$ is not the number of oscillators in the
system which is already infinite from the beginning. To reduce possible
dependences on the value of $L$ we consider periodic boundary conditions
$f_{L+1}=f_{-L}^*,f_{-L-1}=f_{L}^*$ and do the same for the $g$'s. In
our numerical solution we have taken $L=100$ and we have checked that
the results are the same by including more terms in the
hierarchy. Equations have been solved using a second order Euler
algorithm.  We have also started from an initial condition of the type
shown in eq.(\ref{eq7}) with $\theta=0$ in order to check they are not
stationary solutions. Depending on the values of the parameters $K_0$
and $T$ there are different regimes \cite{Bns}. 

In figures 2,3,4 we show the trajectory of the system in the plane
$(Re(f_1),Im(g_1))$ for three different regimes (the incoherent, the
critical and the coherent regimes). Note that according to
eq.(\ref{eq10}) all the trajectories depart from the $Im(g_1)$ in the
direction $-i=\exp(\frac{3\pi}{2})$. The first regime (fig.2)
corresponds to the region where the incoherent solution is stable. In
this case the order parameter $r$ ($r=(f_1 f_1^*)^{\frac{1}{2}}$)
oscillates with an amplitude which decays to zero exponentially in time.
The second regime is shown in figure 3 and corresponds to the the
critical boundary line $T=\frac{K_0}{4}$ where the incoherent solution
becomes unstable. In this case $r$ oscillates and its amplitude decays
to zero algebraically like $t^{-\frac{1}{2}}$ as expected for mean-field
models at the critical point. In the region $T<\frac{K_0}{4}$ the
incoherent solution is unstable and the system reaches a oscillating
stationary solution (see figure 4) in agreement with the results
analytically found by Bonilla et al \cite{Bns,Bcs}. 
Note that this type
of solutions cannot be computed from any theory based on an effective
Hamiltonian (EH) like that given by eq.(\ref{eq6}).  Therefore, the
assumption of the existence of an effective Hamiltonian not only implies
to consider solutions which are not stationary but also to miss another
set of them that are explicitly time dependent.

\begin{figure}
\centerline{\epsfxsize=8cm\epsffile{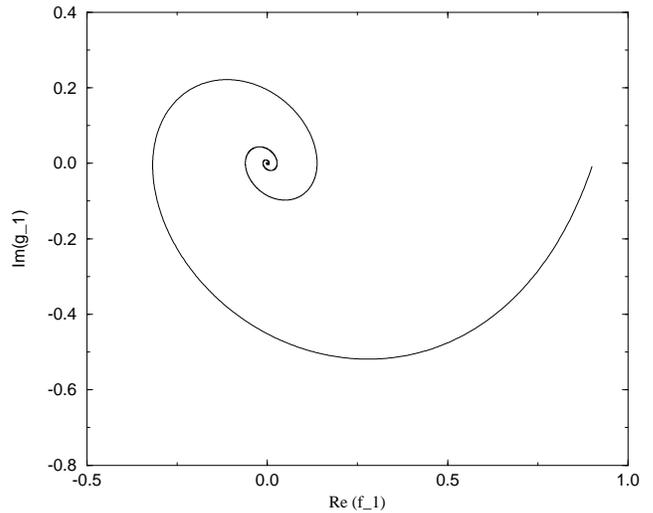}}

\caption{Trajectory of the system in the $(Re(f_1),Im(g_1))$ plane 
at $T=0.5$, $K_0=1$ in the incoherent regime.}
\end{figure}

\begin{figure}
\centerline{\epsfxsize=8cm\epsffile{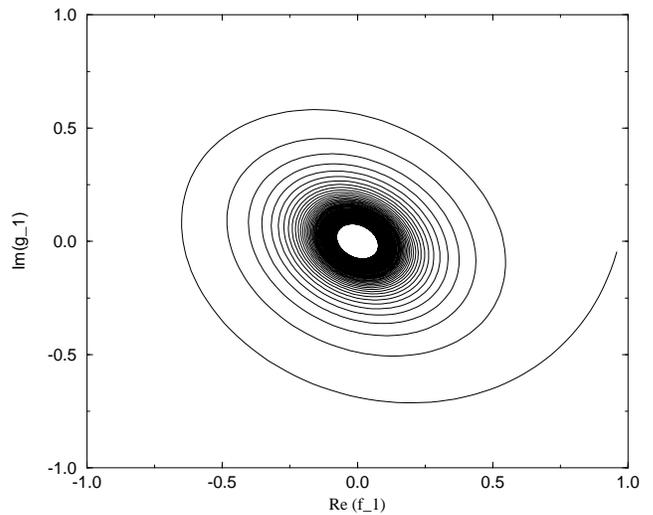}}

\caption{The same as in figure 2 at $T=0.25$, $K_0=1$ in the critical
boundary. The area of the central hole decreases like $1/t$.}
\end{figure}

 We have also observed
the existence of stationary fixed points for enough large values of
$K_0$ as expected when the ferromagnetic coupling is strong enough. In
this case the equilibrium solution within the EH approach albeit
incorrect is closer to the true stationary one (in the limit
$K_0\to\infty$ the EH approach is recovered). 

\begin{figure}
\centerline{\epsfxsize=8cm\epsffile{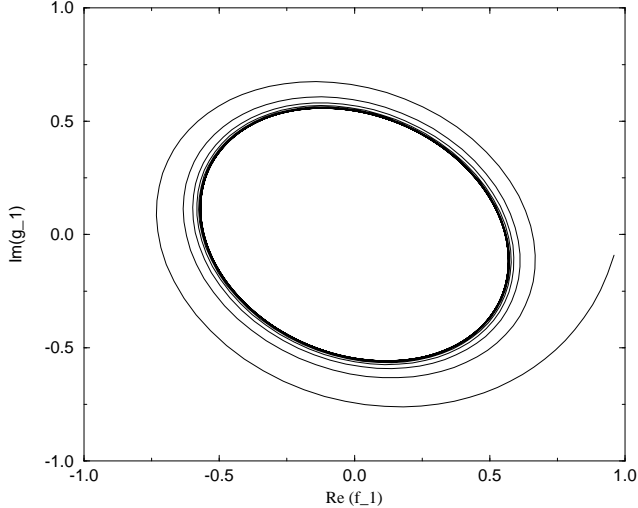}}

\caption{The same as in figure 2 at $T=0.05$, $K_0=1$ in the synchronised
regime.}
\end{figure}

Finally, a comparison
between the present theory (eqs.(\ref{eq11a},\ref{eq11b}) for the bimodal
case) and the Brownian simulations is shown in figure 5. Simulations
have been performed by solving eq.(1) with an Euler method with a time
step $\delta t= 0.005$ and for a population of $N$=50.000
oscillators. Despite some small differences there is remarkable
agreement between the simulation and analytical results.

In summary, we have presented an approach to the analytical solution of
the Kuramoto model which is simple in its formulation and suitable for
analytic and numerical computations. We have shown that the EH approach
fails in predicting the stationary states of the system as well as the
ground states of the energy function eq.(\ref{eq6}). We can give a an
explanation for this result. It has been suggested \cite{vh} that if a
minimum of the energy function eq.(\ref{eq6}) can be localized in the
interior of the region $[-\pi,\pi)$ then this should be asymptotically
stable. In the Kuramoto model it seems that this condition is indeed not
satisfied, at least for the ground states of eq.(\ref{eq6}). As shown in
eq.(\ref{eq10}) the ground states of eq.(\ref{eq6}) are not stationary
states of the dynamics. The quantitative violation of the stationarity
property has been also analitically computed in eq.(\ref{eq10}). The
reason for the discrepancy of our results with those reported in
\cite{vh} relies on how the order of the limits $N\to\infty$ and
$t\to\infty$ are taken. 

\begin{figure}
\centerline{\epsfxsize=8cm\epsffile{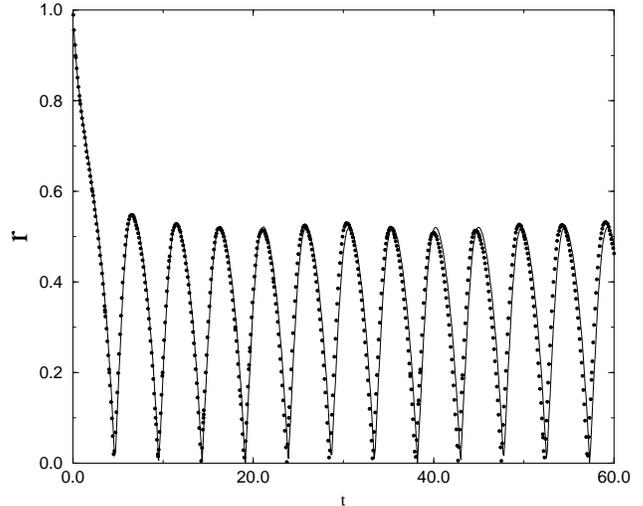}}

\caption{Analytical solution (lines) versus Brownian simulations
(points) in the oscillationg regime with parameters $T=2.5, K_0=1/4$.
Brownian simulations were performed with $50000$ oscillators and one
realization of the noise.}  
\end{figure}

The limit $t\to\infty$ concerns particularly to
the properties of the stationary states. When the limit $N\to\infty$ is
taken first then all ground states of ${\cal H}_{eff}$ become
dynamically unstable. The reason is that in this limit a finite density
of oscillators in the ground state touch the boundaries where the
effective Hamiltonian eq.(\ref{eq6}) is discontinuous. Obviously this is
not true if the limit $t\to\infty$ is taken first. Then, for finite $N$,
the phases $\theta_i$ of the oscillators do not touch the borders
(because it has zero measure) and the ground states are stationary. As said
before, the order of limits (first $N\to\infty$ and later $t\to\infty$)
is the one expected to describe the relevant asymptotic dynamics.  We have
used equations (\ref{eq3}) to investigate the dynamical behavior of the
Kuramoto model free of finite-size effects. Particular results have been
obtained for the bimodal distribution but the method can be generally
applied to any other distribution. It would be very interesting to use
this approach to study the spectrum of correlation and response
functions of the model as well as to investigate the presence of other
dynamical regimes.

{\bf Acknowledgments.}  F.R is grateful to the Foundation for
Fundamental Research of Matter (FOM) in The Netherlands for financial
support through contract number FOM-67596. The work by C.J.P.V has been
partially supported by DGYCIT under grant PB94-0897. We greatly
acknowledge stimulating discussions with A. Arenas,
L. L. Bonilla, A. Diaz-Guilera, F. G.  Padilla and J. M. Rub{\'{\i}} in
the early stages of this work.

\end{document}